\documentclass[
twocolumn,
superscriptaddress,
nofootinbib,
 amsmath,amssymb,
prstper,
floatfix,
]{revtex4-2}

\usepackage{graphicx}
\usepackage{dcolumn}
\usepackage{bm}
\usepackage{color,soul}

\begin{document}

\preprint{APS/123-QED}

\title{Relative impacts of different grade-scales on student success in introductory physics}




\author{David J. Webb}
\affiliation{Department of Physics, University of California - Davis}
\author{Cassandra A. Paul}
\affiliation{Department of Physics \& Astronomy - Science Education Program - San Jose State University}
\author{Mary K. Chessey}
\affiliation{Department of Physics,University of Maryland - College Park}

\date{\today}

\begin{abstract}
In deciding on a student's grade in a class, an instructor generally needs to combine many individual grading judgments into one overall judgment.  Two relatively common numerical scales used to specify individual grades are the 4-point scale (where each whole number 0-4 corresponds to a letter grade) and the percent scale (where letter grades A through D are uniformly distributed in the top 40\% of the scale). This paper uses grading data from a single series of courses offered over a period of 10 years to show that the grade distributions emerging from these two grade scales differed in many ways from each other.  Evidence suggests that the differences are due more to the grade scale than to either the students or the instructors. One major difference is that the fraction of students given grades less than C$-$ was over 5 times larger when instructors used the percent scale.  The fact that \textbf{each} instructor who used both grade scales gave more than 4 times as many of these low grades under percent scale grading suggests that the effect is due to the grade scale rather than the instructor.  When the percent scale was first introduced in these courses in 2006, one of the authors of this paper, who is also one of the instructors in this data set, had confidently predicted that any changes in course grading would be negligible.  They were not negligible, even for this instructor.
\end{abstract}

\maketitle


\section{\label{sec:Introduction}Introduction \& Framing}
In higher education, the role of grades is paramount. Students enroll in courses, and the grade that they receive in each course communicates to the institution the degree to which the student was successful. Passing grades indicate that students are successful and they are able to continue on to more advanced courses in the same topical area. Enough poor grades can cause a student to fail a course and, as a function of the individual university, this can in turn affect the student's time-to-degree, their retention in a major, or even in their retention in college itself. Furthermore, college grade point averages (GPAs) are important as graduate schools and professional schools have minimum requirements for applicants, and some employers request GPA and/or transcript information so course grades influence student career opportunities and pathways even after graduation. Because the stakes are high, it's important that educators take care to construct the meaning behind their grades and also to understand any implications of chosen grading techniques or philosophies. Thus, when the authors of this paper discovered an anomaly in fail-rate data, we sought to make sense of it. This paper explains the anomaly, the analysis we went through to make sense of it, and contextualizes these results in existing literature.

\begin{figure*}
\includegraphics[trim=2.3cm 1.8cm 2.6cm 1.8cm,clip=true,width=\linewidth]{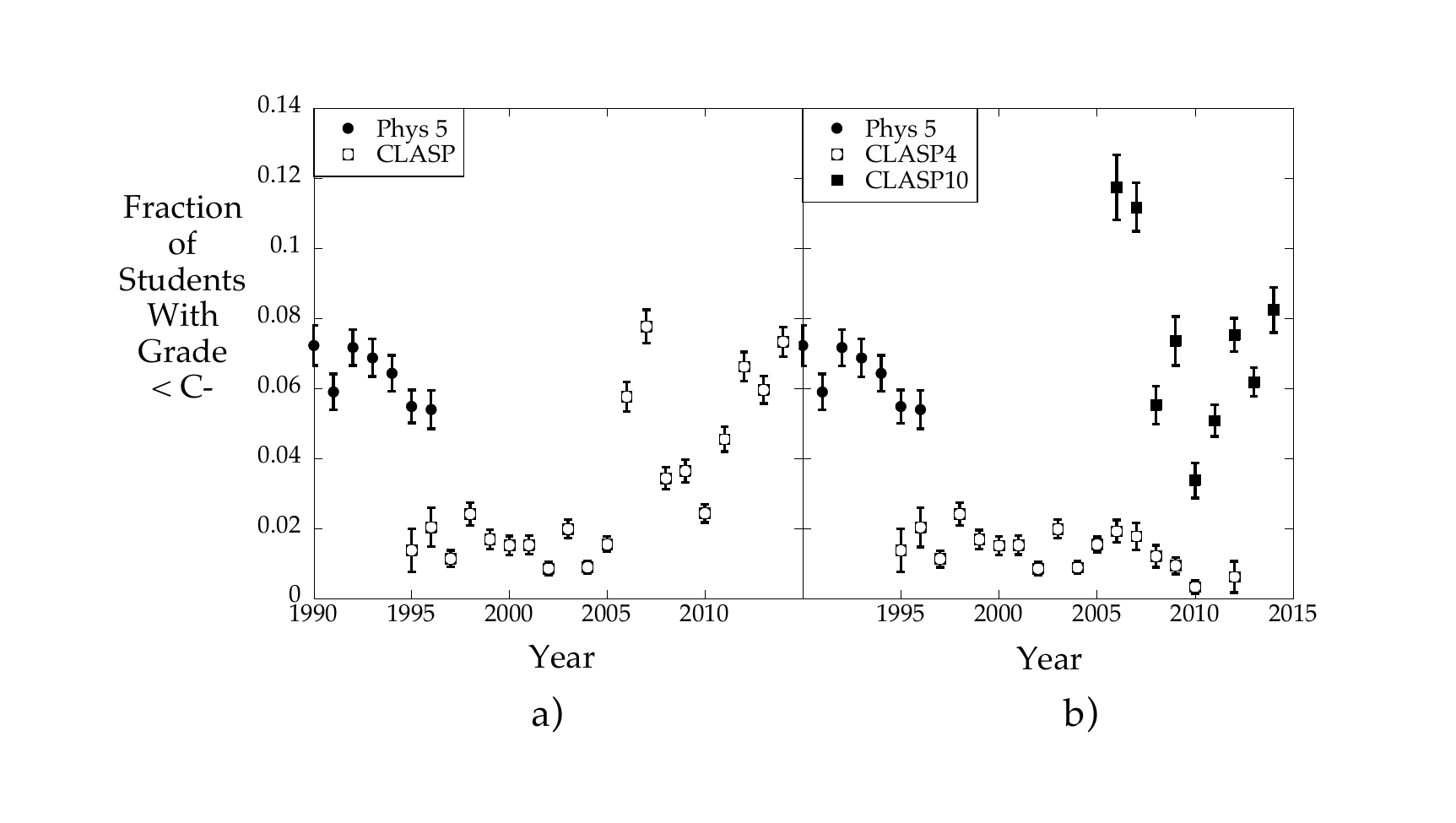}
\caption{Fraction of students that were unsuccessful in physics by year.  The number of students per year in these classes varied from an average of about 2300 in the early 1990's to about 3600 in the middle 2010's.  \textbf{a)} Fraction of students receiving a course grade less than C$-$ as a function of year they took the course.  Phys 5 is the traditional course that was replaced by the CLASP series between 1995 and 1997.  Only the first two courses in the three-quarter series are included for any year because there are many majors requiring only these first two quarters so the student population changes somewhat in the third quarter.  The error bars are $\pm$ standard error.  An increase in the fraction of students with these low grades after 2006 is clear. \textbf{b)} With CLASP4 and CLASP10 grade scales plotted separately, the increase in the fraction of students with these low grades after 2006 can be seen to be due to the use of the 10-point grade scale in many classes.}
\label{Fig0A}
\end{figure*}

In a recent paper \cite{Paul2017} we have compared a traditional physics course at UC Davis to the active-learning based Collaborative Learning through Active Sense-making in Physics (CLASP) curriculum \cite{Potter2014Sixteen} that replaced a more traditional course in 1995.  We showed that students in CLASP courses were significantly more likely to receive a grade $>$D+ than students in the traditional course.  Indeed, students in the CLASP series were more likely to pass the first two parts of this series of courses in exactly two terms and so were less likely to be delayed in their degree progress.  Since publication of this prior work we extended our examination of grades to years well after 1996 and found that in 2006 there was a sudden increase in the fraction of CLASP students receiving grades $<$C$-$ and that this effect continued, at some level, in each year after 2006.  Part a) of Fig. \ref{Fig0A} shows this new result along with the older data.  In trying to account for the change occurring between 2005 and 2006 we examined possibilities such as changes in the student population, changes in administrative rules, changes in the set of teachers teaching the course, and changes in the courses themselves.  We found no significant changes in either the CLASP courses or in administrative rules.  We also checked our other introductory physics series of courses and found no anomalies in fail rate.  We show later in this paper (Sections \ref{sec:PercentScaleFailsMore} and \ref{sec:Selection}) that the other possibilities we mentioned were also relatively unimportant in determining the changes in fail rate that are seen in the figure.  Instead, we found that some of the instructors in the courses had begun, in 2006, using a different numerical grade scale.  After we discovered this, it became obvious that the numerical grade scale an instructor used, mapping letter grades to numbers and vice versa, was a very important factor in this increase in the number of students receiving these low course grades.  The vast majority of CLASP instructors teaching in these years used an absolute grade scale at each level of their grading (exam-item level, exam level, and course grade level), rather than using a curve.  For any specific class, the instructors chose to use one of two grade scales: either a version of the well known 4-point grade scale that we are calling ``CLASP4'' or a 10-point grade scale (CLASP10) that is easily mapped onto another well known grade scale, the ``percent scale''.  Part b) of Fig. \ref{Fig0A} shows that the 10-point scale classes tended to have the highest fraction of students with grades $<$C$-$.  We found that students graded using this 10-point scale were $5.3 \pm 0.4$ times more likely to fail than those graded using the 4-point scale. This difference in fail-rates surprised us, and resulted in further investigation of this phenomena in both our own data set and in existing literature.  

In this paper we examine data from classes that used one of these two grade scales and suggest that, at least for these CLASP courses, the grade scale used for the class was more important in determining the number of students receiving low grades than either the students' incoming grade point average (GPA) or the specific teacher of the class. With this work, we aim to illustrate the large and potentially unintentional impact a choice of grade scale can have on the fail-rate in a large introductory physics course.  

We are not able to make claims on which scale is a more accurate assessment of student understanding or mastery of introductory physics content. However, others have argued that traditional grading methods, including use of the percent scale, are particularly prone to implicit bias \cite{Feldman2018}.  Our prior work finds that some student groups are more likely to be harmed by the percent scale than others. Students who identify as female, first generation, and/or as one of the racial and ethnic groups considered to belong to the category of underrepresented minorities are more likely to leave problems blank on an exam, thus earning zeros for this missing work \cite{paul2018}. Using the common process of averaging grades to determine an aggregate grade results in the percent scale likely being more harmful to these students. In the same work \cite{paul2018} we also show evidence indicating these behaviors are not easily explained as a lack of understanding of physics.  Because we find that using a percent scale in the contexts we are studying has the potential to amplify opportunity gaps, and thus amplify gender, ethnic and racial inequities, \cite{Pell2019} we cannot recommend its usage in contexts similar to the one we describe in this paper, and advise caution in applying it in any context. We share our equity perspective so that readers understand a bias we have against the percent scale, as this bias may sometimes be evident in the paper.

We hope that this paper inspires reflection, discussion, and further research into instructor grading practices.

\section{\label{sec:Background}Background}
While studies examining student scores on standardized concept inventories comprise much of the foundational research  in the field of Physics Education Research (PER)\cite{Hestenes1992, Thornton1998, Hake1998b}, studies examining more typical methods of classroom grading and assessment are comparatively rare. In their review of physics education research field, Docktor and Mestre \cite{Docktor2014} classify physics exams as an area for future study, and state that at the time of their publication in 2014, there is no ``widespread, consistent measure of problem-solving performance.'' On the other hand, specific discussion of classroom grades and grading is absent from their review, which can be interpreted as the topic being at most nominally represented in the field. 

The research in PER that explicitly examines grading practices of open-ended problems on exams finds that instructors use different criteria to decide how many points particular solutions are worth \cite{Henderson2004Grading}, and furthermore that these criteria change depending on what context the work is submitted \cite{Marshman2017}. In both of these studies, the researchers presented graders with example student solutions and asked the instructors to grade as if the students were familiar with their grading practices, implying that the authors have an underlying assumption that instructors will determine grades in a variety of ways that aren't necessarily comparable across different contexts. In these studies, the authors are able to investigate grading practices by comparing grading on two carefully selected example solutions. Both studies find a wide range of given grades comparing across instructors confirming that instructor grading is far from uniform. Because instructors use many different methods to grade, we did not expect one particular change, such as grade-scale, to impact all students across different instructors and courses in roughly the same way.

There is some discussion in the physics education literature describing the process of using different rubrics that are attached to a set of numeric values such as in standards-based grading \cite{Beatty2013, Rundquist2012, Zimmerman2017}, but these papers focus primarily on the process of implementing the grading technique.  There are no papers, to our knowledge, in PER that investigate the impact of grade-scales, or how numbers are assigned and averaged to create a course grade.  On the other hand, the broader field of education research has discussed grading and grade-scales more extensively. 

While there exist many practical guides for classroom assessment and grading (i.e. \cite{Wormeli2006, Guskey2014, feldman2018grading}), grading and grade-scales in the K-12 literature are also understudied. This may be due to the fact that a robust theory for classroom assessment is currently lacking.  Both Brookhart \cite{Brookhart2003} and McMillan \cite{mcmillan2013sage} suggest that classroom assessment theory is emergent and that constructs like validity, reliability, and fairness are as important to classroom assessment as they are to measurement assessment theory, but we have yet to discover how to implement them practically at the classroom level. McMillan claims that more research is needed to inform classroom assessment theory. While theory and research are lacking on classroom grading and grade-scales overall, the issues associated with some aspects of grading are well-documented in the literature. These issues and potential solutions are discussed in the remainder of this section.

\subsection{\label{sec:Percent}Percent Scale}

The percent scale (shown in Table \ref{tab1}), which is also sometimes referred to as percent grade, or standard grade, is a common way of awarding numerical grades to students both in higher education and in the K-12 context in the United States. On the percent scale, 100 is the highest grade students can earn, zero is the lowest, the numbers represent the percent correct, (or more generally, the percent of points earned), and a value around 60-65\% is the boundary between passing and failing (see Table \ref{tab1}) with the location of this boundary depending upon the instructor and the school.  There is a way to interpret this numerical grade as a letter grade that is somewhat standard among nationally recognized institutions (e.g. College Board \cite{collegeboard}, National Center for Education Statistics \cite{nces}). For example, above 90 is generally a grade of `A', between 80 and 90 is usually considered a `B' etc.  Grades below 60 on the percent scale are typically considered failing, and a zero is often awarded for missing work. There are variations of this scale depending on the instructor and/or institutional rules, but we will consider any grade-scale that is based on percent (either out of 10 or out of 100 for example) where students need to earn more than 50\% of the points to earn a passing grade, as percent scale grading. Because instructors take liberties when they implement any grade-scale we refer to these as specific versions of the common percent scale. In Table \ref{tab1} we show the common percent scale, and also share the CLASP10 scale which is a specific version of the common percent scale grading.

\begin{table}[htbp]
\caption{Comparison between different grade scales.  The Letter Grade to Percent Scale and 4.0 Scale conversions are copied from the College Board website. CLASP10 is the specific version of the percent scale whose results are discussed in the paper and CLASP4 is the specific 4-point scale used. (The scales are named ``CLASP'' because they are used in the Collaborative Learning through Active Sense-making in Physics course studied in this paper, which is further discussed in section \ref{sec:Setting}.) }
\label{tab1}
\begin{ruledtabular}
\begin{tabular}{c c c c c }
\textbf{Letter} & \multicolumn{2}{c}{\textbf{Common Scales}} & \multicolumn{2}{c}{\textbf{Specific Scales}} \\
\textbf{Grade} & Percent Scale & 4.0 Scale & CLASP10 & CLASP4 \\ 
 \hline
A+ & 97-100 & 4.0 & 9.67-10 & 4.17-4.5 \\
A & 93-96 & 4.0 & 9.33-9.67 & 3.83-4.17 \\
A$-$ & 90-92 & 3.7 & 9.0-9.33 & 3.5-3.83 \\
B+ & 87-89 & 3.3 & 8.67-9.0 & 3.17-3.5 \\
B & 83-86 & 3.0 & 8.33-8.67 & 2.83-3.17 \\
B$-$ & 80-82 & 2.7 & 8.0-8.33 & 2.5-2.83 \\
C+ & 77-79 & 2.3 & 7.67-8.0 & 2.17-2.5 \\
C & 73-76 & 2.0 & 7.33-7.67 & 1.83-2.17 \\
C$-$ & 70-72 & 1.7 & 7.0-7.33 & 1.5-1.83 \\
D+ & 67-69 & 1.3 & 6.67-7.0 & 1.17-1.5 \\
D & 65-66 & 1.0 & 6.33-6.67 & 0.83-1.17 \\
D$-$ &  &  & 6.0-6.33 & 0.5-0.83 \\
E/F & 0-65 & 0.0 & 0-6.0 & 0-0.5 \\
\end{tabular}
\end{ruledtabular}
\end{table}

\subsection{\label{sec:Criticisms}Criticisms of the Percent Scale}
In recent years, the percent scale has come under criticism \cite{Guskey2013, Reeves2004, Connor2011, Wormeli2006} and alternative methods of grading have been introduced. One criticism of the percent scale is the portion of the scale devoted to failing grades. In a 2013 article, Guskey \cite{Guskey2013} points out that a larger portion of the scale is devoted to failure (65\%) than success (35\%).  This means that failure (grades equivalent to `F') can be measured in 65 different degrees, while each of the other letter grades are limited to only 10 degrees (A, B, C) or 5 degrees (D). Another way of stating this is that the grade space devoted to `F' is roughly 6 times larger than that of any other letter grade. That the majority of the scale is devoted to F is potentially a philosophical problem; in fact Guskey \cite{Guskey2013} asks, ``What message does that communicate to students?'' But the problems are mathematical as well because, as discussed by Connor and Wormeli \cite{Connor2011} any of the F-grades below 50 tend to skew an averaging procedure \cite{Reeves2004, Wormeli2006}.  An example of the ``skewing'' downward in an averaging process is seen in Table \ref{tab2} comparing a 4-point grade scale with a percent scale in averaging the grades on a hypothetical exam.

The amount of the scale devoted to `F' grades is particularly important  when considering awarding the lowest grade, a grade of 0 \cite{Guskey2013, Connor2011, Reeves2004, Wormeli2006, paul2018}.  Zero grades are often given to students who leave an exam answer blank or even skip an assignment altogether. This may be justified as a way to encourage students to do their best to always submit some work.  However, research \cite{Wormeli2006,Selby1992,Grant2003} has shown that while good grades are motivators of good work, poor grades may not always motivate students to work harder.  While there are conflicting viewpoints on whether missed assignments should be included in aggregate grades at all (see Chap's 14-17 of Ref. \cite{Wormeli2006}), if we take the viewpoint that there are some instances where a earning a 0 is warranted, critics point out that an instructor using the percent scale makes it difficult for a student to balance out the effect of even a single zero.  Table \ref{tab2} shows the result of leaving one answer blank on a hypothetical exam consisting of four problems where the student gave good to excellent answers on the first, second, and fourth problems but left the third problem blank.

\begin{table}[htbp]
\caption{Grades a student might earn on an exam with 4 problems when the third problem was left blank.  The resulting exam grade is calculated, by averaging, using either of the two CLASP grade scales (see Table \ref{tab1}).  Leaving an answer blank clearly carries much more weight in averaging under CLASP10}
\label{tab2}
\begin{ruledtabular}
\begin{tabular}{c c c c}
\textbf{Question} &
\textbf{Grade} &\textbf{CLASP4}
&\textbf{CLASP10}\\ 
 \hline
1 & B$+$ & 3.2 & 8.7 \\
2 & A & 3.9 & 9.4 \\
3 & F & 0 & 0 \\
4 & B$-$ & 2.7 & 8.2 \\
 \hline
Exam Grade &  & 2.45 (C+) & 6.58 (D) \\
\end{tabular}
\end{ruledtabular}
\end{table}

Finally there are a number of criticisms of the scale that are not mathematical in nature, for example the fact that low failing grades (grades below 50\% for example) negatively motivate students, or reduce their self-efficacy \cite{Cavallo2004, Hazari2010}.  These are also important arguments to consider, but our quantitative data does not address these issues and so in this paper we are primarily concerning ourselves with the mathematical implications of grade scale.

\section{\label{sec:AltGradeScales}Alternative Grade Scales and Practices}

There are a few different common alternatives to using the percent scale. For example, in order to mitigate the percent scale issue of devastating zeros, the concept of ``minimum grading'' was conceived \cite{Connor2011}.  Minimum grading is the practice of raising very low grades to some `minimum' grade (usually 50\%) so that students are able to recover from missing work, or just really poor assignment performances.  Critics of minimum grading suggest that it may result in students passing even if they have not learned the material. They argue that minimum grading promotes student entitlement (they get something for nothing) and leads to social promotion. Some instructors feel strongly that it's unfair to give students 50\% if they have completed less than 50\% of the work. There are also concerns that minimum grading contributes to well-documented grade inflation \cite{Rojstaczer2012} (a phenomenon that may include a higher average grade, more students being given `A's, or both). Research on minimum grading \cite{Carey2012} in one school district shows that neither of these things happened. In one of the few large-scale quantitative studies on classroom grading, Carey and Carifio \cite{Carey2012} present an analysis of seven years of grading data collected from a school that implemented minimum grading. They used standardized test results to show that students who earned at least one minimum grade actually outperformed their peers who did not receive any minimum grades on standardized testing. This was true even though the students receiving minimum grades had, on average, lower classroom grades. This suggests that minimum grading need not cause grade inflation but also that minimum grading may not entirely make up for the inequities of a percent grading scheme.  Unfortunately, even though this seems to be a reasonable way to address some concerns of the percent scale, many instructors don't like it \cite{Reeves2004}, because they do not agree with the concept of giving students scores that they don't believe they have earned.

Another way to address the percent scale issues is to use the concept of Standards-Based grading (discussed in Brookhart, et al review \cite{Brookhart2016} and references therein). This method asks students to demonstrate proficiency in certain areas, with the instructor providing ordinal grades such as well below proficiency, approaching proficiency, proficient, and excellent, that describe the path to proficiency \cite{Anderson2018}.  This approach has been used \cite{Beatty2013} in college physics classes but, because the method requires students to be able to have multiple opportunities to attempt the same proficiency, it is difficult to accomplish with the large class sizes that are typical of introductory science courses. 

A final alternative \cite{Guskey2013} to the percent scale is the college 4.0 scale which is typically used to calculate GPA.  Each integer in the scale (4, 3, 2, 1, 0) corresponds to a letter grade (A, B, C, D, F). While many college level instructors may use the percent scale and then convert to a letter grade (which has a numeric value tied to the 4.0 scale), others may use a 4.0 scale from the beginning of the course and then report the numbers in the form of letter grades.  These numbers can be averaged into a single grade. Because the 4.0 scale allocates the same `space' to each letter grade, the scale avoids most of the issues associated with the percent scale. For example, in Table \ref{tab2} we saw that a zero carries much less weight in an averaging procedure using a 4.0 scale than in averaging under a percent scale . This means that a grade of `zero' is less disastrous for students. This scale is mathematically similar to the practice of minimum grading, and instructors using standards-based grading also sometimes use the integers on a 4.0 scale to mean different levels of proficiency \cite{Anderson2018, Beatty2013}, so aspects of this scale exist in both of the other alternatives we describe. Because the 4.0 scale mitigates many of the criticisms of the percent-scale, many suggest it as an alternative to percent scale grading. 

\section{\label{sec:Questions}Research Questions}
The active learning introductory physics course CLASP at UC Davis, has been in existence since 1995 \cite{Potter2014Sixteen}. Originally, all instructor teaching the the CLASP courses used the same grade scale described as ``CLASP4'' in Table \ref{tab1}. This grade scale was based on the standard college 4.0 scale. After many years, instructors began to move away from the CLASP4 grade scale, and began to utilize a percent scale instead. Because the course materials over the years were extremely similar, and some instructors used both types of grade scale in different sections of the same course, the circumstances provide for an ideal opportunity to compare the usage of examples of the two scales. In this paper we consider CLASP10 to be an example usage of the percent scale, and CLASP4 to be an example of a 4-point scale. We use these data to compare use of the percent scale, CLASP10 to the 4-point scale, CLASP4, to examine some common critiques of the percent scale and to further explore similarities and differences between these two grade scales. 

\textbf{Specifically, regarding the controversies over percent scale grading, we ask:}

1) Does the 4-point scale lead to course grades that are ``inflated'' compared to percent scale grading?

2) How does the distribution of student course grades differ between the percent scale and 4-point grading? 

3) How does the distribution of exam-item grades differ between the percent scale and 4-point grading? 

4) How variable are the course grades for each grade scale by instructor and by class? 

When considering these questions, it's important to emphasize that our aim here is to uncover how different grade scales can impact student outcomes. We are not evaluating the philosophy behind either scale, nor making claims about what constitutes an ``A'' or an ``F.''  Finally, we will make no claims about the connection between assigned grade and student understanding. 

In this article we examine 10 years of student grades in these CLASP courses.  The ten years 2003-2012 bracket the introduction of the CLASP10 grade scale in 2006.  What we will show is that the fraction of students failing the course is much larger when instructors use the percent scale (CLASP10) and that instructors assigned more F grades to their student's work when using the percent scale.  We also find that the increase in students failing is associated mostly with the grade scale used when aggregating grades and not with the increase in individual assigned F's.  The grade scale is the most important thing because the presence of the very low F-grades in a percent scale can skew course grades much lower, in the averaging procedure, than can F-grades in a 4-point grade scale.  These conclusions also seem to be more related to the grade scale than to the instructor. Finally, we find that that using the percent scale leads to more class-to-class variability in grade distribution.

\section{\label{sec:Methods}Methods}

\subsection{\label{sec:Setting}Setting and Context}

The grading we discuss took place in introductory physics classes that were designed for biological science and agricultural science students attending UC Davis, an R1 research university.  Over the years discussed in this paper the average high school GPA of freshmen varied from around 3.7 in the earlier years to around 3.9 in the later years.  Over these years the average SAT total varied from high 1100's in the early years to mid 1200's in later years and admission-offer rates varied from around 70\% of applicants in the earlier years to of order 50\% of applicants in the later years. 

When calculating a GPA, most colleges use a 4.0 scale which equates each integer (4, 3, 2, 1, 0) to a corresponding letter grade (A, B, C, D, F). The designers of the Collaborative Learning Through Active Sense-making in Physics (CLASP) \cite{Potter2014Sixteen} curriculum wanted a grading system that was both transparent and non-competitive, and so they directly linked every single graded item (be it an exam question, or the exam itself) to a slightly modified version of this 4-point scale so that students could understand how their performance on a given question related to the expectations of the course instructors. The resulting CLASP4 grade scale, shown in Table \ref{tab1}, is therefore a version of the 4.0 scale. CLASP4 is a continuous grade scale from 0 to 4.5 with each letter grade region, except for the F region, centered on the appropriate integer. Generally, a grade of 4.5 (the highest A+) was earned by a student when their description/calculation correctly and completely applied an appropriate physical model to the situation described in the exam.
Student exam descriptions/calculations that were not correct or were not complete were assigned lower grades with each grade depending on the instructor's judgment of the quality of the answer. The graders used a scoring method called ``Grading by Response Category'' (GRC) \cite{Paul2013Grading} in which a grader would categorize student responses by their most significant error, and then give the same score and written feedback to all students who made that error.  This type of scoring cannot be considered a rubric because the categories are made after looking at student responses, but are otherwise similar to holistic rubrics \cite{Brookhart2004} in that scoring is subjective (requires judgement as the answer isn't simply correct or incorrect) and that a single score and feedback is given for each exam problem. An answer judged to be excellent but not perfect was given a score in the A$-$ to A$+$ range between 3.5 and 4.5, an answer judged to be good was given a score in the B$-$ to B$+$ range between 2.5 and 3.5, an answer judged to be satisfactory was given a score in the C$-$ to C$+$ range between 1.5 to 2.5, an answer judged to be unsatisfactory was given a score in the D$-$ to D$+$ range between 0.5 to 1.5 Under this grade scale a zero was almost universally reserved for students who did not answer the question at all, the top of the F range of grades was 0.5, and grades between 0 and 0.5 were used for students who gave an answer but whose answer showed almost no familiarity with or understanding of the subject. Multi-part exam problems often had a grade for each part (exam-item level grades) and these were averaged, with the weight per part determined by the instructor, to determine the exam grade. These exam grades were then averaged, with weight per exam determined by the instructor, to give the course grade on the same 4-point scale.  From 1995 until 2005 essentially all CLASP instructors used this same basic grading method for quizzes, exams, and the class grades\footnote{Homework was generally not graded and there are no homework grades in these databases but see Section \ref{sec:NonExamGrades} for a discussion of the effects of various non-exam grades.}

In 2006 several instructors began experimenting with a 10-point grade scale, CLASP10, that was just a re-scaled version of the standard percent scale.  With the CLASP10 grade scale, answers in the A$-$ to A$+$ range were given grades 9 to 10 (instead of the 3.5 to 4.5 of the 4-point scale), B$-$ to B$+$ range were given grades 8 to 9, etc. for C and D grade ranges.  Again the zero of this grade scale was reserved for students who did not answer the question but now the highest F grade given is 6.0.  The result is that, relative to CLASP4 scale, CLASP10 had a much larger grading measure available for F's (0-6) even though the other grades have the same measure on each scale.  For these reasons we will often refer to CLASP10 as a ``percent scale''.  The grades on a single exam problem were then, as under CLASP4, averaged to give the exam grade and the exams were similarly averaged to give a course score which then largely determined the course letter grade.

\subsection{\label{sec:Data Set}Data Set}

Data were collected from the first two quarters of the three quarter CLASP series from course archives spanning the ten years 2003-2012.  The ``CLASP A'' content primarily covers energy and thermodynamics, while the ``CLASP B'' content focuses on mechanics.  During those years the structure and content of these classes was relatively constant.  Over 75\% of a student's time in one of these classes was spent working in discussion/laboratory sections (referred to as DLs in the CLASP curriculum) on activities that only changed slowly over those years.

The platform for entering grades in these large enrollment classes was centralized using a separate database file for every separate course offering.  These course databases include exam scores for each student along with the individual grades that were given as well as the calculations that led to these exam scores. In addition, these databases sometimes included the calculations that led to the actual course grades. 

Over these ten years there were 133 of these classes, and we have found databases for 95 of them that are identifiably graded as described above using either CLASP4 grading or percent scale grading. This identification was determined by examining the maximum grades given to individual student answers. If the maximum was always 10 then we considered the class to have had percent grading and if the maximum grades were always 4.5 then we considered the class to have had CLASP4 scale. The average class size is about 250 students and was about the same for each grade scale.  The resulting database contains 773,667 exam-item level grades given to 15,757 individual students on each part of each exam.  Fifty-six of the included classes are CLASP4 grade scale (including 478,617 grades on individual answers) and the remaining 39 are percent grade scale (including 295,050 grades on individual answers).

We also have access, from UC Davis administration, to the recorded course grades from all of these classes as recorded by the Registrar and to each student's GPA upon entering the CLASP course. Course grades were all recorded on the standard 4-point grade scale with letter grades A through D, each of which may include a $+$ or $-$, and F.  In each database that included the calculation of course (letter) grades these letter grades were determined using the cutoffs approximately as shown in Table \ref{tab1}.  Although we will group together the data from the first two courses (CLASP A and CLASP B) in this series of courses because the differences between the grade scales show up in both courses, we will specifically note each situation where the data from one course differ substantially from that of the other.

Each of our research questions requires a different set of comparisons to make. So rather than providing a list of justifications for the comparability of each set in this section, we instead share this information when it can be considered alongside the research question and resulting comparison.

\section{\label{sec:Results}Analysis \& Results}

Throughout our analysis we compare the percent scale, CLASP10, to the 4-point scale, CLASP4, used by the instructors of CLASP. In addition to course grades we will also be reporting on individual exam-item scores that instructors gave to student answers on exams. These scores will generally be referenced by the grade range in which they are contained. For instance, a 3.8 given to a student answer under CLASP4 grading will be considered to be in the A region just the same as a 9.3 given to a student answer under CLASP10 grading.  We treat a borderline grade (i.e. a 3.5 under CLASP4 grading or the equivalent 9.0 under CLASP10 grading) given to a student as being half of a grade in each of the bordering letter grade ranges because we are focused on the averages and a borderline grade contributes exactly the same to an average regardless of whether it is considered part of the upper range or part of the lower range.  We will point out if and when varying that choice for borderline grades makes a difference in our conclusions.  Most counts, averages, standard errors, etc. were calculated in Excel and most are double-checked with STATA software. We used STATA for the standard statistical tests, calculations, and regressions. The error estimates we give will be standard error of the mean or propagated from standard errors unless otherwise noted.

\subsection{\label{sec:PercentScaleFailsMore}Percent scale fails more students}
\begin{figure}
\includegraphics[trim=3.7cm 2.9cm 5.0cm 3.5cm, clip=true,width=\linewidth]{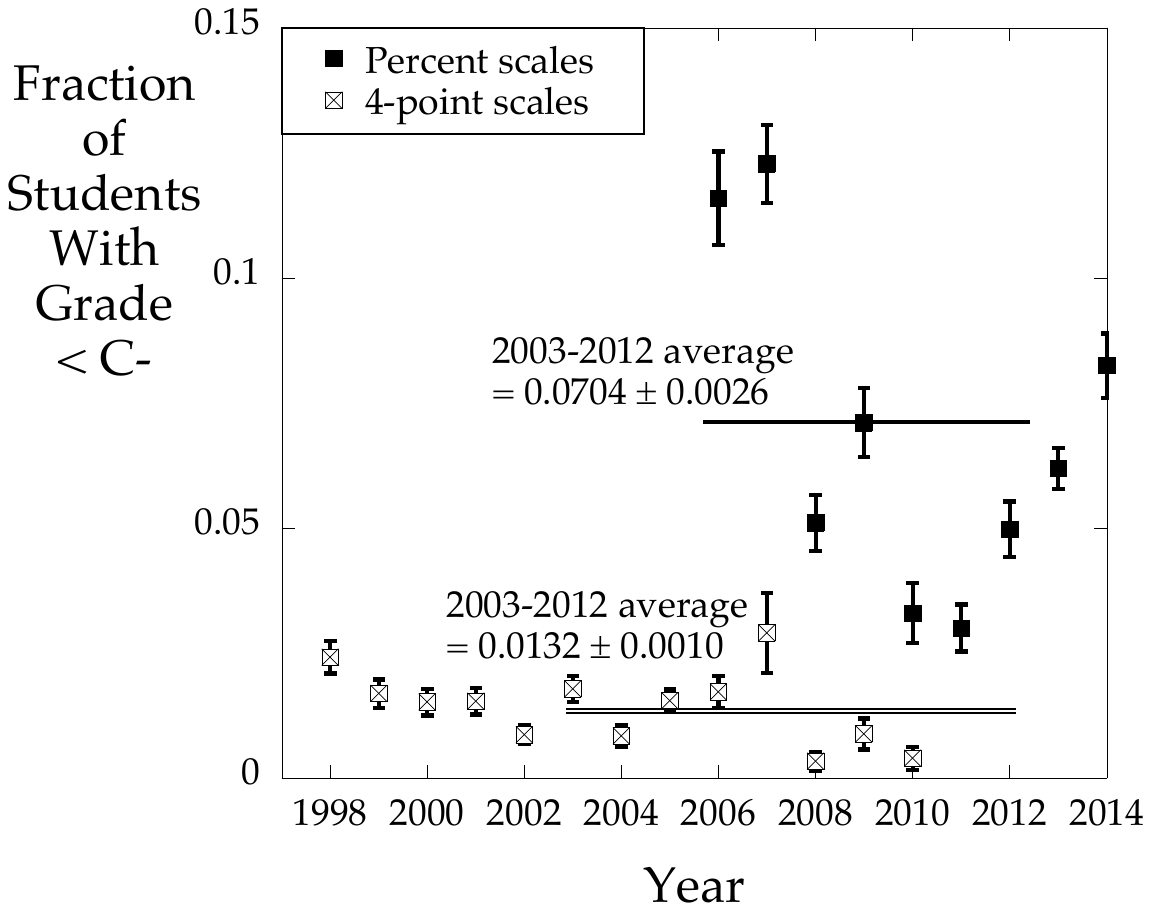}
\caption{Fraction of students receiving a course grade less than C$-$ as a function of year they took the course.  The two main grade scales used in the course are shown separately.  The error bars are $\pm$ standard error.}
\label{Fig1}
\end{figure}

When one separates the classes taught according to grade scale, either 4-point or percent scale, a trend in student fail-rates is undeniably present. Figure \ref{Fig1} shows the fraction of course grades given that are less than C$-$ as a function of the year.  We choose a cutoff of C$-$ to measure because UC Davis allows students receiving less than a C$-$ to repeat a course whereas those with C$-$ or higher cannot.  A similar way of noting that C$-$ is an important cutoff is that courses graded Pass/Not-Pass give the Pass grade only to students who would have received a grade of C$-$ or higher. Therefore, the figure shows the fraction of students who have a grade considered low enough to warrant repeating the course. From the averages shown on the figure we see that instructors using the percent grading scale gave $5.3 \pm 0.4$ times as many grades lower than C$-$ than instructors using a 4-point scale.  This fraction was $4.4\pm 0.5$ in the first course in the series (CLASP A) and $7.0\pm1.0$ in the second course (CLASP B) so the two courses both showed this grade scale effect.  A chi-square test for the CLASP A data set shows this difference between grade scales is significant, $\chi^2(2, N=13,249) = 247.6, P < 0.001$ and the same test for CLASP B is also significant, $\chi^2(2, N=10,078) = 279.9, P < 0.001$. Of course, it's conceivable that these differences are primarily an issue of student academic performance rather than of grade scale used. To control for academic performance we use i) the variable $EnterGPA$, a student's GPA upon entering the course, as a predictor of that student's academic performance along with ii) the variable $GrScale$ (the grade scale) as a categorical variable in a regression model.  We use logistic regression because the data meet the requirements of this method but, because the data are heteroskedastic, don't meet the requirements for multivariable linear regression.  The model we use in predicting the odds of a student receiving a grade less than C$-$ including both of these independent variables is 
\begin{multline}
ln[odds(CourseGrade < C-)]\\=b_0 + b_{GPA} EnterGPA + b_{GrScl} GrScale   \label{equ:OddsLTCminus}
\end{multline}
where $e^{b}$ is the appropriate odds ratio (note that the odds of receiving a grade less than C$-$ is equal to the probability of receiving a grade less than C$-$ divided by the probability of receiving a grade of C$-$ or higher).  For CLASP A classes this model gives us an odds ratio for $EnterGPA$ of $0.074 \pm 0.010$ and we find that a student graded under a percent scale had $5.5 \pm 0.7$ times higher odds of receiving less than C$-$ than the same student under 4-point grading ($z=14, N = 11,804, P<0.001$ for the variable $GrScale$ in this model) after controlling for entering GPA.  Similarly, for CLASP B classes this model gives us an odds ratio for $EnterGPA$ of $0.087 \pm 0.013$ and we find that a student graded under a percent scale had $8.1 \pm 1.2$ times higher odds of receiving less than C$-$ than the same student under 4-point grading ($z=14, N = 9,546, P<0.001$ for the variable $GrScale$ in this model) after controlling for entering GPA.  We conclude that student academic performance does not explain the large fraction of students with these low grades under percent grading.  Finally, we should point out that the student withdrawal/drop rates are essentially independent of grade scale ($0.76\%\pm0.07\%$ under 4-point grading and $0.74\%\pm0.09\%$ under percent grading) in this set of classes.  What isn't shown in these data is why more students were failed when the percent scale was used. In the remainder of this paper we analyze factors that contribute to this phenomenon.

\subsection{\label{sec:GradeInflation}Is Grade Inflation Happening?}

Because instructors select either the percent scale or the 4-point scale, it's important to consider the instructors also in this analysis. One possibility is that instructors using the 4-point scale give generally higher grades to their students' answers than those given by instructors using a percent scale, a situation commonly referred to as ``Grade Inflation'' when it leads to an overall increase in high grades.  If 4-point scale instructors were simply inflating grades, we would expect to find fewer low grades but we would also expect to find more high grades under 4-point grading as well as a higher average grade.

\begin{figure}
\includegraphics[trim=3.5cm 3.2cm 5.4cm 2.9cm, clip=true,width=\linewidth]{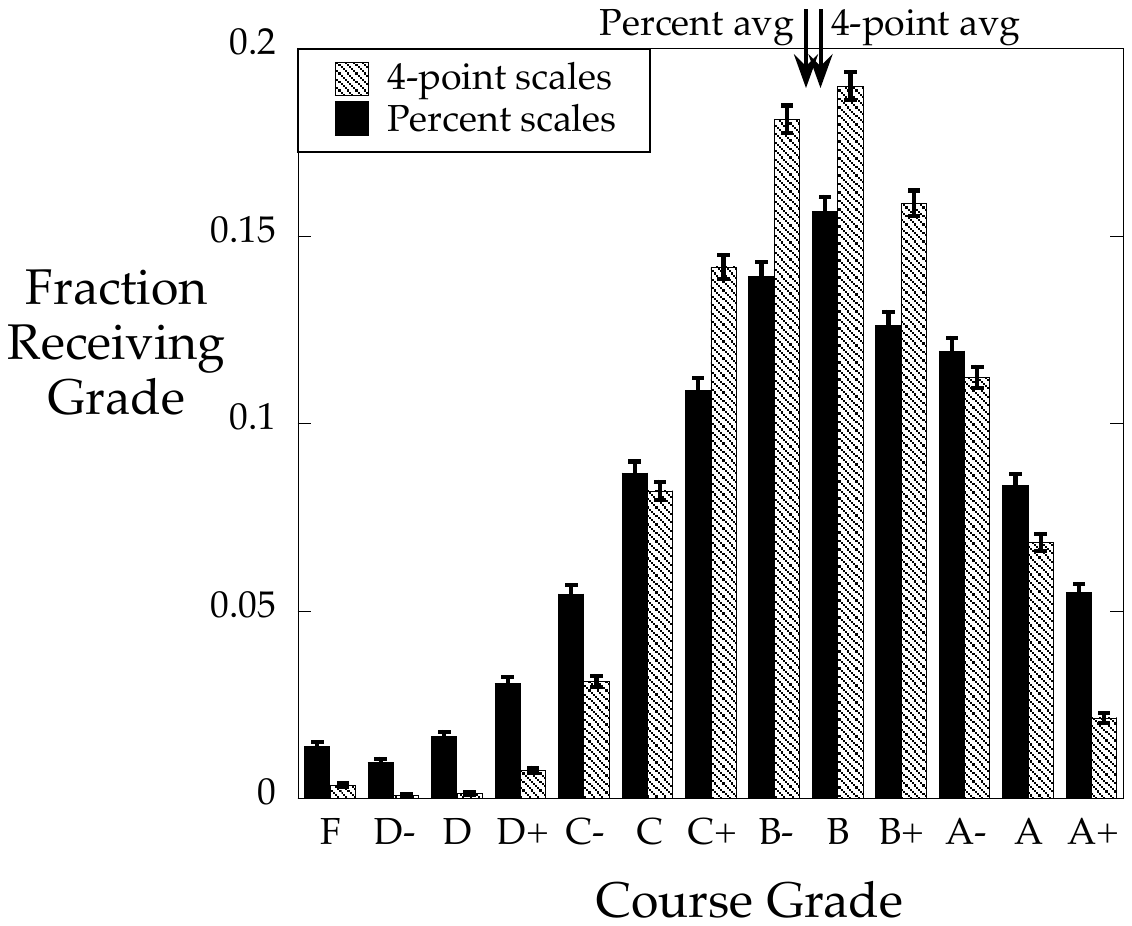}
\caption{Fraction of course grades given between 2003 and 2012 for each of the two major grade scales.  Error bars are $\pm$ standard error.  The average grade is between B$-$ and B for each grade scale but the distribution for the 4-scale classes is notably narrower.}
\label{Fig2}
\end{figure}

Figure \ref{Fig2} shows the complete course grade distribution for both grade scales.  One possible sign of grade inflation in 4-point classes, more high grades, is easily seen to be missing.  Instead, we find that students in percent scale classes are the ones receiving more A's (about 20\% more than their peers in 4-point classes).

To look for a shift in the average grade, we computed the average course grade given by each grade scale using the UC Davis method for calculating GPA (A=4.0, A$-$=3.7, B+=3.3, B=3.0, etc.) except that we use A+=4.3 rather than the UC Davis A+=4.0. This amounts to choosing, for each course grade, a value roughly in the middle of the relevant CLASP4 range shown in Table \ref{tab1}.  We find the average grade given to a student graded under the percent scale was 2.852 (SD = 0.89) and under the 4-point scale was 2.918 (SD = 0.67) for a grade shift of 0.066 (0.01). These average grades (both between a B and a B$-$) are shown in Fig. \ref{Fig2}.  The effect size, using Cohen's d, of this grade shift is about 0.086 so it is a small effect.  The difference of 0.07 GPA units is less than half of the class-to-class variation for either grade scale (standard deviation, over the individual classes, of average class grade is 0.19 for 4-point classes and 0.34 for percent scale classes).  The two courses that we have grouped together here showed different results for this comparison when considered individually.  The first-quarter course (CLASP A) had essentially the same average grades for the two grade scales ($2.884\pm0.008$ for 4-point grading and $2.899\pm0.012$ for percent grading).  A t-test of the two CLASP A distributions shows that this small difference in the two average grades is not statistically significant ($t=1.1, df=13,247, P=0.288$) and including student GPA as a covariate does not change this conclusion. The second quarter course (CLASP B) had a lower average under percent grading ($2.961\pm0.009$ for 4-point grading and $2.787\pm0.014$ for percent grading).  A t-test of these two CLASP B distributions shows that this difference in the two average grades is statistically significant ($t=11.3, df=10,076, P<0.001$) and including student GPA as a covariate does not change this conclusion.  So there is conflicting evidence for any simply defined ``grade inflation.'' We find that students graded using the percent scale are more likely than students graded using the 4-point scale to earn ``A'' grades, but that they are also more likely to fail the class.  When we examine the distribution of grades for both courses in figure \ref{Fig2}, we see evidence that suggests fewer students fail under the 4-point scale not because the distribution of course grades under that grade scale is shifted uniformly toward higher grades but that the course grade distribution under 4-point grading was narrower than it was for percent scale grading.

\subsection{\label{sec:UseofGradeSpace}Instructor Use of Grade Space}

Since very low grades given under percent scale grading have a much larger effect on course grade than the lowest grades given under 4-point scale grading (Table \ref{tab2}), we might expect that this is the main difference between the two grade scales and accounts for the difference in the student fail-rates.  However, there is another clear difference in our data that can also lead to more failing grades.  We find differences in how instructors allocate grades on the individual answers given by their students on exams that differ between the two scales.

\begin{figure}
\includegraphics[trim=3.6cm 3.1cm 5.7cm 3.1cm, clip=true,width=\linewidth]{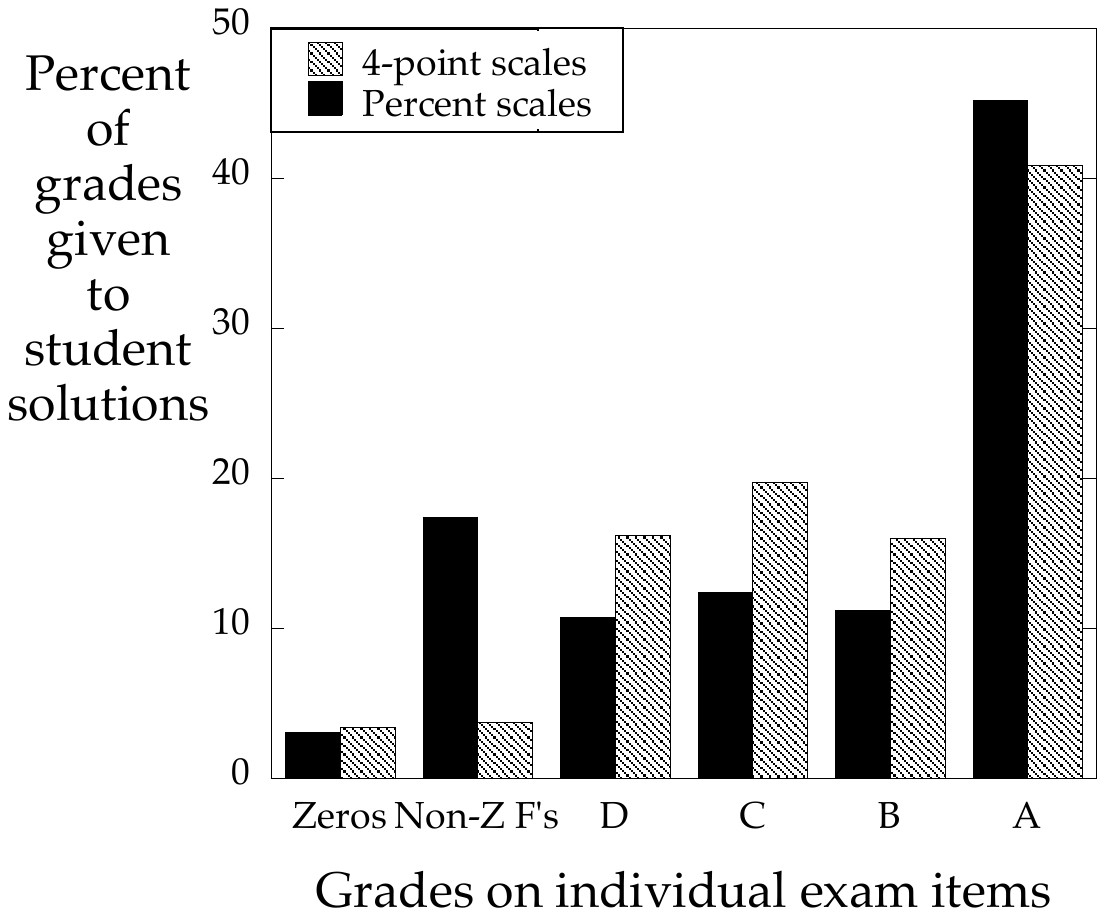}
\caption{Fractions of grades given to student answers on individual exam problems under each of the two grading regimes, 4-point scale and percent scale. The standard errors are quite small (range from $\pm$ 0.03\% to $\pm$ 0.12\%) and so are not shown.  Nevertheless, the choice of how to distribute the grades assigned at the boundaries does affect the totals.  The largest effect is for non-zero F's where percent scales have an extra weight (above 4-point grading) of between 13.3\% (border grades counted in higher category) and 14.4\% (border grades counted in lower category) .}
\label{Fig3}
\end{figure}

First, we note that students who leave a problem blank (or mostly blank) on an exam receive a zero, a grade that involves no instructor judgment of understanding or skill.  Figure \ref{Fig3} shows the fractions of scores given on individual exam items for all 96 courses.  We see that there is very little difference between the number of zeros given to students under each scale. However, the fraction of F-grades that are not zero is distinctly dependent on the grade scale. Figure \ref{Fig3} shows that instructors using a percent scale are considerably more likely to judge individual student solutions on exam items as non-zero F's than those instructors using the 4-point scale.  This amounts to shifting about 14\% of the entire exam item grade weight from higher grades under 4-point grading down to non-zero F's under percent grading.  Notice that none of this shifted grade weight comes from the A's, which were actually more common under percent grading and likely little of it from B's because the total fraction of (A's + B's) is about the same for the two scales, 56.4\% for percent grading and 56.9\% for 4-point grading.

In order to analyze how these grade shifts affected individual students and whether they might be due to student academic performance, we compute the fraction of non-zero F's for each student.\footnote{In determining a student's fraction of non-zero F's, a grade on the borderline between D and F is equal to half of a non-zero F-grade.}  Averaging this student-level number over all students gives $0.0364\pm 0.0004$ under 4-point grading and $0.1628\pm 0.0014$ under percent grading.  The two courses that we have grouped together here had different amounts of grade weight shifted into non-zero F's.  In the first course of the series (CLASP A) $7.5\%\pm0.8\%$ of grade weight was shifted into nonzero F's under percent grading and in the second course of the series (CLASP B) that number was $18.5\%\pm0.2\%$.  The student-level distribution of non-zero F's is, unfortunately, both non-normal and heteroskedastic but we can still use student GPA, $EnterGPA$, in a linear regression to see if it affects the extra fraction of non-zero F's seen for percent grading.  The regression model we use to model a student's fraction of non-zero F's is
\begin{multline}
FractionOfNonZeroFs\\=b_0 + b_{GPA} EnterGPA + b_{GrScl} GrScale
\label{eqn:NonZeroFs}
\end{multline}
For CLASP A classes this model predicts that the percent scales had $8.6\%\pm0.1\%$ of extra grade weight in non-zero F's compared to 4-point scales, $t=61, P<0.001$ for the variable $GrScale$, and $N=11,804$ and adjusted $R^2=0.30$ for this model.  For CLASP B classes the model predicts that the percent scales had $18.0\%\pm0.2\%$ of extra grade weight in non-zero F's compared to 4-point scales, $t=86, P<0.001$ for the variable $GrScale$, and $N=9,546$ and adjusted $R^2=0.46$ for this model.  These two results both suggest that controlling for student GPA does not change the fraction of shifted grade weight very much so we conclude that the extra non-zero F's under percent scale grading are likely not due to differences in the students in these courses.

\subsection{\label{sec:GradePrecision}Instuctor Use of Indivdual Grades}

Over 99.6\% of the exam item grades given included at most one decimal place (i.e. a grade of 8.8 was relatively common but a grade of 8.75 was not).  This would seem to imply that CLASP10 supplies instructors with a finer grading scale to use because there are about twice as many possible grades.  However, in a practical sense the two grade scales used about the same number of distinct grades.  For instance 11 distinct CLASP10 grades make up 80\% of the CLASP10 grades given but it takes at least 13 distinct CLASP4 grades to make up 80\% of the grades under CLASP4 so CLASP4 instructors are using more distinct grades more often (80\% is just an example but any percentage up to 98.5\% works as well).  In total, the 46 distinct CLASP4 grades make up 100\% of the grades given and the same number of CLASP10 grades make up 99.5\% of the CLASP10 grades given.  This means that half of the CLASP10 grades available to the instructors were used, in total, for only about 0.5\% of the grades given under that grade scale.  Twenty-two of the extra CLASP10 grades were not given even once among the 478,617 grades.  We find that, extra grade space afforded by CLASP10 was used primarily to give integer-valued low-F grades and was not used by the instructor for finer-scale grading.

\subsection{\label{sec:NonExamGrades}Effects of non-exam grades}
Since the beginning of CLASP in 1995, course grades have standardly included the possibility of a grade change from the bare exam grade with these grade changes being based on the student's work in their discussion/lab (DL) section. These grade changes were almost always grade increases and would commonly be given to 10-30\% of the students. This DL grade could be based solely on participation but, since students go over their homework in discussion/lab, it could also include their homework in some way.  In addition to these DL grades, some instructors would ask ``clicker questions'' in lecture and the students' answers to these questions would be folded into the grade in some way that also generally resulted in an increase over the bare exam grade.

We expect that our comparison between CLASP4 and CLASP10 is dominated by the grade scale used to grade exams.  As a check we estimate the effects of non-exam grades.  To do this we used all of the class databases for which we had i) each student's total exam grade based on either the 4-point scale or the 10-point scale and ii) the total course grade number that was used to determine the students' course letter grades using the appropriate grade scale. Instructors calculated this total course grade by folding together the exam grade, DL grade, and clicker score in some way, so the difference between this course grade and the total exam grade measures the effect of these extra grades on the course grade. We have 46 CLASP4 class databases that include the data we need and 20 CLASP10 databases that include these data. Using these databases we find that the average grade increase for a particular class ranged from 0.002 points to 0.20 points over these 46 CLASP4 classes with an overall student-level average increase of 0.039 (grade) points.  In comparison the CLASP10 classes had class-average grade increases ranging from -0.004 points to 0.56 points with an overall student-level average increase of 0.17 points. 

Because the CLASP10 course grades that we have full access to are, on average, slightly inflated by these non-exam-based grades compared to the CLASP4 classes, leaving these grades out (i.e. using only exam grades) would likely have led to an even bigger difference in the fail rates of these two grade scales. Nevertheless, in keeping with the general culture of the CLASP courses, the average instructor for either grade scale based most of the course grade on their student's exam performance.

\subsection{\label{sec:Selection}Individual Instructors' Results}
One possible explanation for the difference between the failing rates of the two grade scales is that different instructors choose the scale that serves their interest better. Therefore, it is useful to address any selection effects the choice of grade scale may have. Of the 60 instructors involved in these courses over the 10 years in our data set, seven instructors used both the 4-point scale and the percent scale at various times. This gives us seven comparisons between the two grade scales where an instructor\footnote{Each non-summer course has two instructors, both of whom are responsible for the grades given.} is held constant.  Table \ref{tab3} shows that these seven instructors gave between 4 and 12 times more course grades less than C$-$ under percent grading than under 4-point grading and between 2 and 11 times as many non-zero F's on individual exam items.  These numbers are modified only very slightly if we attempt to account for differences in the student academic performance by using their incoming GPA as a covariate as in Equations \ref{equ:OddsLTCminus} and \ref{eqn:NonZeroFs} but the conclusions stated above do not change.  These data, taken together with the data of Sections \ref{sec:PercentScaleFailsMore} and \ref{sec:UseofGradeSpace}, are evidence against either instructor or students as causes of the effects shown in Fig's \ref{Fig1}, \ref{Fig2}, and \ref{Fig3}.

\begin{table}[htbp]
\caption{Seven instructors used both grade scales at various times. N's are the numbers of students taught under the particular grading regime.  The Fail Ratio is the fraction of that instructor's students with course grades less than C$-$ under percent grading divided by that fraction under 4-point grading.  Similarly, the Non-Zero F Ratio is the fraction of that instructor's exam-item grades that were non-zero F's under percent grading divided by that fraction under 4-point grading.  Standard errors are in parentheses.}
\label{tab3}
\begin{ruledtabular}
\begin{tabular}{c c c c c c}
\textbf{Instr.} & \textbf{N} &\textbf{N} & \textbf{Fail}
& \textbf{Non-Zero F}\\ 
\ & \textbf{CLASP4} &\textbf{CLASP10} & \textbf{Ratio}
& \textbf{Ratio} \\
\hline
1 & 2869 & 1741 & 7.4 (2.0) & 4.8 (0.1) \\
2 & 1508 & 975 & 7.8 (1.7) & 3.8 (0.1) \\
3 & 1959 & 271 & 4.1 (1.1) & 2.1 (0.1) \\
4 & 603 & 520 & 4.2 (1.2) & 3.7 (0.1) \\
5 & 3633 & 772 & 4.6 (0.9) & 7.4 (0.2) \\
6 & 676 & 302 & 7.8 (6.3) & 11.1 (1.2) \\
7 & 1299 & 263 & 12 (4) & 7.4 (0.3) \\
\end{tabular}
\end{ruledtabular}
\end{table}

\subsection{\label{sec:Variability}Variability of the Percent Scale}
\begin{figure*}
\includegraphics[trim=1.7cm 4.7cm 1.6cm 4.7cm,clip=true,width=\linewidth]{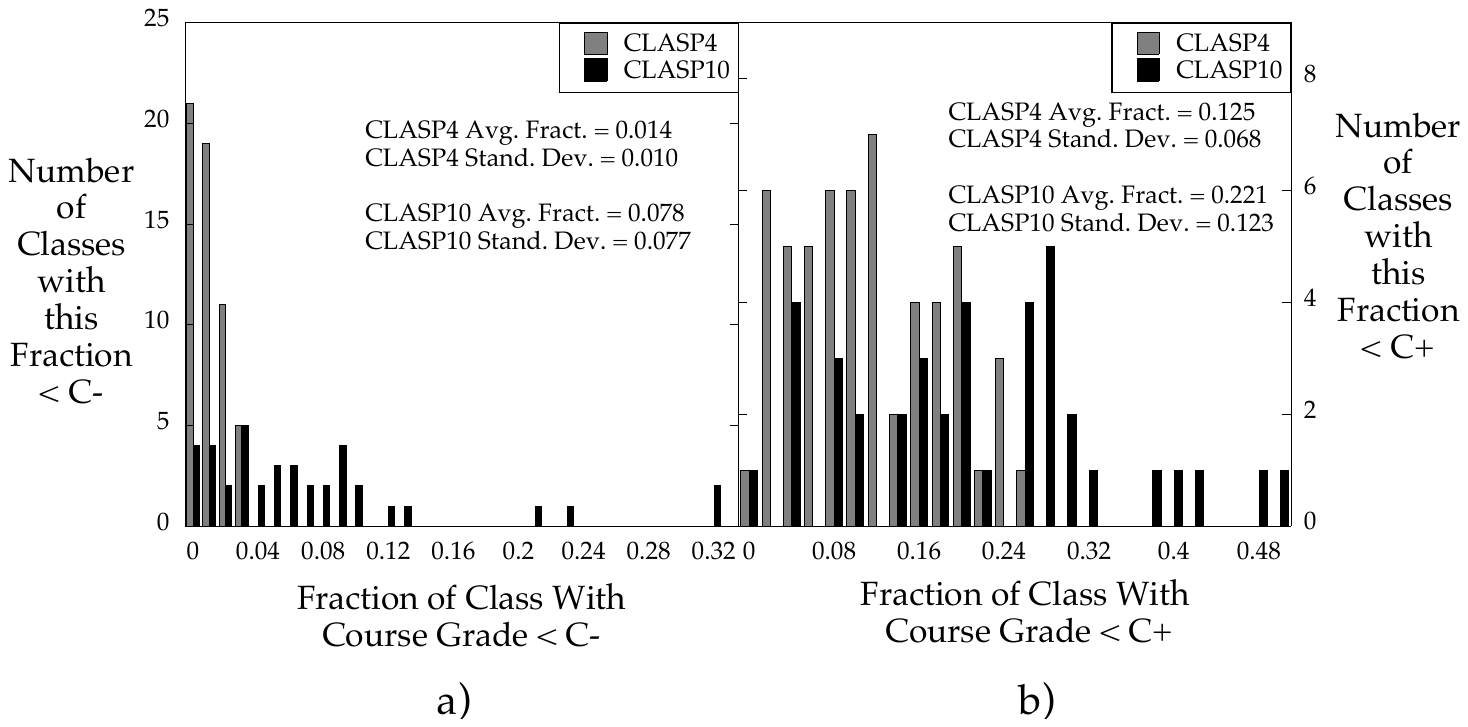}
\caption{\textbf{a)} Histogram showing how classes (categorized by grade scale CLASP4 or CLASP10) are distributed with respect to the fraction of students with course grade $<$ C$-$.  CLASP10 graded classes had a much larger variation in this fraction, extending from 0 to over 0.3.  \textbf{b)} Same as part \textbf{a)} except the class fractions are for students with course grade $<$ C$+$.  The higher cutoff is used to remove possible ``floor'' effects (both grade scales had classes in the lowest bin but neither had a class with zero grades less than C+).  The variation of this fraction is also larger for CLASP10 than for CLASP4.}
\label{Fig5}
\end{figure*}

Figure \ref{Fig1} shows that the fail-rates for the instructors using the percent scale are more variable by year than the 4-point scale. We can quantify this on a class-by-class basis by examining how the fail rate (i.e. fraction of students with grade $<$ C$-$) varies over the classes offered over the years included in this study.  Figure \ref{Fig5}a) is a histogram showing how the class-failure-rate is distributed over the classes in our data, for each grade scale. One finds not only that the failure rate is larger for the percent scale courses but that it is also much more variable over courses.

Because both of the distributions in Fig. \ref{Fig5}a) include classes at the lower bound of zero students failing, we might worry that being pushed against this lower bound has artificially narrowed one or both of these distributions, a ``floor'' effect.  We check if this is influencing our conclusion by using a higher grade cutoff.  A cutoff of C still had classes from each scale with zero students below cutoff, so we use a cutoff of C+.  Figure \ref{Fig5}b) shows the distribution over classes of the fraction of students with course grades less than this higher cutoff. The percent scale classes are more variable under this measure also. 

Some might argue that the coefficient of variance (COV = SD/mean) is an appropriate metric when comparing variations for distributions with different means. By this metric we find that the class-to-class coefficient of variance for CLASP10 is larger when considering the fraction of grades less than C$-$, Fig. \ref{Fig5} a), but about the same for fractions of grades less than C$+$, Fig. \ref{Fig5} b).  However, focusing our attention on the relative variation (the COV) rather than the absolute variation might lead to minimizing the very real impacts on the students of the variations in this fail-rate.  Under CLASP10 grading a student enrolling in a class may have a 2\% chance of getting a grade less than C$-$ or they might have a 30\% chance.  Under CLASP4 that probability varies only a little from the average of 1.4\%. Thus we claim that there is a large and meaningful difference, in course to course variation, between the scales.  

In the Appendix we use student GPAs to show that this class-to-class variability (for either grade scale) has little or nothing to do with any obvious class-to-class variations in the students' academic abilities.  We also show that the larger variation within CLASP10 classes, compared to CLASP4 classes, combines both an intra-instructor increased variation and an inter-instructor increased variation.

\subsection{\label{sec:Weight vs. F}Effects of the Percent Scales' Low-F's}

Table \ref{tab2} showed that a very low F grade (for example, 0/10) has a much larger effect on the average grade under percent scale grading than any F-grade under 4-point scale grading.  On the other hand, the highest percent scale F grades (for example a 6/10) average in much the same way that F's do with a 4-point scale.  Since the lower F-grades require more excellent grades to cancel them out, we can think of them as F's that carry more effective ``weight'' than any F from a 4-point scale.  These ``weighty'' F's lead to the ``skewing'' of average grades discussed by Connor and Wormeli \cite{Connor2011}.

As noted earlier, some of the class database files that we have access to not only have all exam-item grades recorded but also include all of the calculations that led to the exam grades and eventually to the course grades.  Courses offered in 2008 and later years do not have all of these calculations but eight of the thirteen CLASP10 graded classes offered 2006-2007 do have complete sets of exam-item grades and all of the grade calculations including the final course grade. These eight classes gave 1839 course grades to 1272 students (567 students are in two of these particular classes). The eight classes give us a way to separate the effects of the heavy weighing of low F's from the effects of just giving more F's on the individual exam answers. Without changing the number of F's (or any other grade) that were given on exam-items in these eight classes we map the CLASP10 percent scale grades onto the CLASP4 scale as follows: i) for all exam-item grades larger than 6.0 we subtract 5.5 from the grade and ii) in order to be very conservative in our treatment of the F's we set all original CLASP10 exam-item grades less than or equal to 6.0 to a 0 in our 4-point re-scaling so that all of the F's are given the highest weight possible under CLASP4. We then do all the original weighted averages of exam-item grades into exam grades and exam grades into course grades where point cutoffs determining the letter grades are gotten by subtracting 5.5 from all of the original percent scale grade cutoffs. The original percent graded courses had a fail rate of 8.3\% $\pm$ 0.6\% and the 4-point re-scaled courses had a fail rate of 1.3\% $\pm$ 0.3\%.  The 4-point rescaled fail rate of 1.3\% is consistent with the rest of the 4-point graded classes (see Figures \ref{Fig1} and \ref{Fig5}) and the increase of a factor of 6.5 $\pm$ 1.4 in the fail rate under percent grading is consistent with both the overall ratio of 5.3 and the individual instructor ratios given in Table \ref{tab3}.  These overall consistencies from a simple rescaling onto a 4-point scale suggests that \textbf{the heavy effective weight that the low-F grades carry is the main factor increasing the failure rates} and that the extra F's assigned by instructors using the percent scale for individual exam answers aren't the main difference between the two course grade distributions.

\section{\label{sec:Discussion}Discussion}

Considering only the mathematical characteristics of each scale, it is perhaps not surprising that more students fail under the percent scale. After all, if lower grades are given more weight, it's likely that more students will fail. That said, there are several nuances in the overall grade distributions of each scale uncovered by our analysis that are worth considering by instructors who are considering using either scale.

The above results confirm many of the critiques of the percent scale discussed in Section \ref{sec:Criticisms}.  A complete understanding of the grade scale entails understanding these critiques so that the percent scale can be used consistently and effectively. In fact, many instructors are aware of some of these critiques, and adjust percent scale grades accordingly (for example, they might consider a lower grade than 65\% passing or they might add some number of points to all students' grades to increase the class average.) The intention of this discussion is to discuss the nuances of the use of the percent scale in one particular context in order to bring these characteristics attention so that they may be discussed both by the research community and instructors considering their grading philosophy.

\subsection{\label{sec:partial credit} Considering Partial Credit}
If a student does not complete a problem correctly but shows some small part of understanding, instructors will often award an accordingly small amount of ``partial credit.'' Figure \ref{Fig3} shows that a very large portion of grades earned on exam items graded using the percent scale is devoted to ``Non-Zero F's.''  In fact, more students earn ``Non-Zero F's'' on the percent scale than any other individual grade besides ``A''. Mathematically, there is a big difference between averaging in a 40\% and a 10\%, but does the instructor see a correspondingly meaningful difference between these two different grades for failing?  Perhaps, but is this the difference between these two failing grades equivalent to distinguishing the differences between a grade of D (65\%) and an A (95\%) which are also 30 points apart?  If an instructor thinks that the numerical grades they assign are actually interval in nature, rather than just ordinal \cite{Stevens1946,Brookhart2004}, and they are averaging the individual grades to determine some aggregate grade, then the same consideration as to whether to award a D or an A should go into determining whether a student earns an F (10\%) or an F (40\%). 
Specifically, it seems important that instructors avoid the mind frame of awarding percents lower than 50\% and thinking that this is giving the student partial credit if they consider a grade of roughly 60\% to be the border between passing and failing. For example, while 30\% is obviously better than a zero it is still 30 points below failing, which is the same mathematical difference between an A and an F.  The data in our paper show the collective effects of the very low F-grades of the percent scale.  And the fact that each instructor who used both grade scales gave more F's suggests that the grade scale itself might affect a teacher's grading judgments in cases when a student's answer does not show much understanding.

\subsection{\label{sec:Considering zeros} The Meaning of a Zero}
The effect of awarding a zero is greater in the percent scale than in the 4-point scale, as shown by Table \ref{tab2} and therefore contribute to the grade weighting issues discussed in section \ref{sec:Weight vs. F}.  When a student leaves an exam problem blank for any reason, the instructor often awards them a zero for this. The instructor's justification for this is certainly logical in the sense that the student has provided zero evidence of understanding, and therefore has earned 0\% of the possible points. The instructor might also be using the zero for a motivational purpose in the sense that students need to complete the work in order to earn points. However, many studies \cite{Grant2003,Shim2005,Corpus2009} suggest that this simple view of motivation is unwarranted. In addition, as seen in Fig. \ref{Fig3}, our data set shows that even though a zero carries much more weight on the percent scale than on the 4-point scale, the overall fraction of blank problems remains essentially constant. This is tentative evidence in support of the fact that the number of blank responses is not affected by choice of grade scale in this course and possible additional evidence against the motivational use of a zero.  Furthermore, our prior work has shown that leaving a physics problem blank is a behavioral trait that is more common  for women, students identifying as underrepresented minorities, and first generation college students in this context \cite{paul2018}. Therefore the practice of awarding zeros for missing work may well contribute to achievement and opportunity gaps.

As pointed out by other authors \cite{Guskey2002,Wormeli2006}, a zero that is earned because a student did not complete an assignment is not a measurement, it's actually missing data. In previous work \cite{paul2018} we showed that the number of problems left blank by a student is poorly correlated with other metrics of understanding, so that leaving a blank is by no means predictive of that student's overall understanding of physics.  Averaging in zeros for missing data would be a terrible practice in one's research and so we might consider alternatives to this practice should this concern us in our teaching.

\subsection{\label{sec:Valuing the Instructor's Evaluation}Valuing the Instructor's Evaluation}

In Fig. \ref{Fig3} we showed that the instructors using the percent scale gave more non-zero ``F'' grades than instructors using the 4-point scale and, using Table \ref{tab3}, argued that these extra F's were due to the grade scale rather than the instructor. However later, in Section \ref{sec:Weight vs. F}, we argued that the reason that more students failed the course when graded using the percent scale as compared to the 4-point scale, was due to the mathematics of averaging percent scale grades, and NOT because those instructors actually gave more ``F'' grades on individual exam items when using the percent scale.  In some ways this second point seems to temper the first point but the two effects should really be considered together.  For example, regarding the second point, an instructor could argue that giving a very low F grade and using it in an averaging process that gives it a large effective weight is entirely appropriate because that is what their student ``earned.''  However the first point, that the percent scale seems to have guided seven out of seven instructors into giving more of these F grades, might give this instructor pause regarding their own ability to provide impartial absolute grading judgments of these poor student answers.  The set of low F grades which average with higher effective weight are exactly the grades that may be caused by the grade scale itself and so be particularly hard to be confident about.

\subsection{\label{sec:Considering Variation at the course level}Considering Variation at the course level}
In Fig. \ref{Fig5} we show that there is a larger range in fail-rates for CLASP10 courses than for CLASP4 courses. Variation of fail-rates at the course level in itself is not problematic. Should students in one class learn more, or otherwise show that they met the course goals more completely than those in another course, a difference in average course grades or fail-rates would be expected and warranted. We can not rule out this possibility entirely. However, the courses studied here used the same curriculum, the same pedagogy, and very similar assessment formats. They are also all large introductory courses ranging from 70 to 300 students, with the average course size of 250.  Given the large sizes of these classes it is not surprising that the students in CLASP4 classes were indistinguishable from those in CLASP10 classes in terms of their GPA distributions, see Appendix \ref{sec:AppendixA}.  With so much being the same, and with a large number of students in each course, we would not expect a big variation across courses, and yet we find that the standard deviation of fail-rates in CLASP10 courses is much larger than that of CLASP4 courses, see Section \ref{sec:Variability}.  This indicates to us that the likelihood of a student passing a course is dependent on which section they enroll in CLASP10 courses to a much greater extent than it is for CLASP4 courses. This finding is potentially of interest to institutions trying to increase equity across different sections of the same course. And the 4.0 scale could be a viable alternative to curving grades to a specific distribution. (See Chapter 4 of Guskey, 2014 \cite{Guskey2014} for discussion on curving.)

\subsection{\label{sec:Limitations}Limitations \& Future Work}
These results are derived from a data set that comes entirely from two (sequential) courses offered over a period of ten years by one department at a single institution. This course has a fairly low fail rate regardless of which scale is used. The course grades we discuss in this paper are largely determined using a weighted average of exam grades and that grades due to non-exam parts of the course were much less important. On average  the non-exam parts of the course modified the exam grade by no more than a fifth of a grade point no matter which grade scale was used.  The pencil and paper exams were graded using the Grading By Response Category \cite{Paul2013Grading} method of subjective scoring.  Findings would likely vary across institutions and different course contexts where pencil and paper exams hold less of the grading weight.  While we have made every effort to account for student and instructor selection effects, this is not a randomized controlled study.  It is possible that there are unseen factors contributing to higher fail-rates in the courses graded using the percent scale. 

Future work will examine similar data sets at other institutions offering CLASP and other Introductory Physics for Life Sciences courses.  Furthermore, a big question we have not addressed here is what happens to these students after completing this course.  Are students who would have failed under the percent scale, but passed under the 4-point scale successful in future courses?  We do not address this question in this paper, as our intent is not to prove one grading method is superior to another, but rather to uncover characteristics of the percent and 4-point scales that are important for instructors to consider when deciding on a grading practice for their courses. We do plan to investigate this question in forthcoming work.

\section{\label{sec:Conclusions}Conclusions}

In summarizing our results we again emphasize that we are not casting judgment on use of the percent scale in general or any grading practices in particular, but instead argue that it is essential for instructors to consider the biases of the percent scale when planning their course for the semester, so that they can ensure that their teaching philosophies match their grading philosophies. 

The primary purpose of these analyses is to fully understand the impact of the percent scale as compared to another somewhat commonly used scale, the 4-point scale. In fact, even knowing these results, one of the authors has chosen to continue using a version of the percent scale in their small graduate courses. In this use case, work that does not meet expectations earns no lower than 60\%, but zeros for missing assignments are used to ensure it's not possible to pass the course without completing the assignments (students in this example have the option to submit late work without penalty). This example is shared NOT as an example of an exemplar use of the percent scale (we do not have such data to support such a claim), but rather to emphasize that the intention of this paper is not to discredit the percent scale, but rather to expose some characteristics of the percent scale that may have previously avoided consideration due to its widespread use.

 With respect to our research questions, we draw the following conclusions.

RQ1) The 4-point scale does not appear to inflate grades in the traditional use of the word because the average grades of the two scales are close and students are actually over 30\% \textbf{less likely} to earn A grades in courses using the 4-point scale.

RQ2) Although the average course grades under the two scales are close, the width of the distribution is much larger under percent grading. This led to many more students receiving failing course grades when the percent scale is used as compared to the 4-point scale.  We found that the odds of failing is over 5 times higher under percent scale grading, $P<0.001$.  The overall fail rate varies by instructor and by individual class for the percent scale with an average fail rate of about 8\%. Nevertheless, \textbf{each} instructor who used both scales at various times failed, overall, at least four times as many students under the percent scale.

RQ3) Instructors tend to give out more ``F'' grades on individual exam items when using the percent scale than when using the 4-point scale (13\% to 14\% of the entire grade weight was shifted down into the F-region under percent grading, effect size = 1.3).  We have not seen this sort of effect reported in the literature. However, the number of extra F's is not found to be the main contributor to the higher course fail rate under percent scales. Rather, the extra ``effective weight'' of these low F's in the averaging process is the main contributor. The grade scale is more important in determining the fail rate than the instructor.

RQ4) The percent scale has a much more variable fail rate when compared to the 4-point scale. The class-to-class variation of the fail rate under percent grading is over seven times higher than under 4-point scale grading even though the variation in the students was negligible. We show that a larger variation under percent grading seems to be independent of any floor effect in Fig. \ref{Fig5}.

Each college or university level instructor has their own opinion about the quality of a student's work but this judgment should represent an unbiased opinion of that work.  Toward that end, it is useful for instructors to know the origins of possible biases so that they can account for these in assigning grades. Our results indicate that instructor use of the 4-point scale led to many more students passing their introductory physics course as compared with classes using the percent scale. This result was achieved without grade inflation. Our findings align with previous critiques of the percent scale, and indicate that instructors should consider the specific issues we highlight in this paper when using the percent scale.

\section{\label{sec:Acknowledgements}Acknowledgements}
None of these results would have been possible without the organized databases that Wendell Potter set up in 1997 and that continued past his retirement in 2006 so, in memoriam, we owe him a debt of gratitude. We would also like to thank the education research groups at UC Davis and San Jose State for useful comments on the research and the manuscript.  We also would like to thank Jayson Nissen for providing feedback on an earlier draft of this paper.

\appendix
\section{\label{sec:AppendixA}ANALYSIS OF CLASS-TO-CLASS VARIATION}
Figure \ref{Fig5} showed that the fraction of students in a class who received low grades was much more variable under CLASP10 grading than under CLASP4 grading.  It is possible that the increased variability under percent scale grading happened because the students themselves were more variable in percent scale courses.  We can check this using the students' incoming GPA's to find the distribution, over classes, of the class-average GPA ($AvgGPA$), for each grade scale.  We find that the average over CLASP4 classes of $AvgGPA$ is 2.97 with a standard deviation over classes of 0.08 and we find that the average over CLASP10 classes of $AvgGPA$ is 2.99 with a standard deviation of 0.08.  A t-test comparing $AvgGPA$ for the two grade scales gives $t=1.2, df=93, P=0.24$ so the classes under the two grade scales are, in terms of $AvgGPA$, indistinguishable from each other.  We can also examine whether the fraction of students in a class who receive a grade $<$C$-$, $FracFail$, varies with the average incoming GPA, $AvgGPA$, of that class.  The regression model we use for this is
\begin{equation}
FracFail = b_0 + b_{AvgGPA} AvgGPA
\end{equation}
where our dataset includes 56 CLASP4 classes and 39 CLASP10 classes.  For the two grade scales we find the following regression coefficients: $b_{AvgGPA} = -0.012 \pm 0.016$, P = 0.49, $R^2=0.009$ for 4-point scales and $b_{AvgGPA} = -0.20 \pm 0.15$, P = 0.19 $R^2=0.05$ for percent scales.  The large P-values and small values of $R^2$ suggest no significant dependence of a class's $FracFail$ upon that class's $AvgGPA$.  A possible problem with this analysis is that the class-averaged GPA might not contain enough information about how many low GPA students (those most likely to receive grades $<$C$-$) are in any particular class.  We can calculate the fraction ($LowGPAFrac$) of students who have incoming GPAs below some cutoff and use as that fraction as a covariate in the linear regression.  We note that more than 80\% of students getting grades less than C$-$ come into the class with GPAs less than 2.8 so we use that cutoff to define $LowGPAFrac$ and examine the linear dependence of $FracFail$ upon this $LowGPAFrac$.  Our model for this regression is
\begin{equation}
FracFail = b_0 + b_{LowGPA} LowGPAFrac
\label{eqn:FailvsGPAFrac}
\end{equation}
For the two grade scales we find the following regression coefficients: $b_{LowGPA} = 0.027 \pm 0.019$, P = 0.164, $R^2=0.018$ for 4-point scales and $b_{LowGPA} = 0.21 \pm 0.16$, P = 0.20 $R^2=0.043$ for percent scales.  Again, the large P-value and small value of $R^2$ suggests that the variations of an instructor's students are not obviously giving rise the the variations in the fail rate.  These general conclusions are robust under changing the incoming GPA cutoff to 2.4, 2.6, or 3.0.  So our general conclusion is that the variability in failing fraction under CLASP10 grading is not obviously associated with an underlying variability in the students' expected academic performance.  We remind the reader that both the incoming GPA of an individual student and the grade scale their instructor uses are significant predictors (see Section \ref{sec:PercentScaleFailsMore}) of the odds of that student failing the course.  However, the variability of this fail rate is apparently related only to the grade scale and not to any obvious variability in the students.

If the variability in fail-rates is not associated with the students and it is not obviously a floor effect then maybe it is an inter-instructor effect where any specific instructor fails about the same fraction of students in each percent scale course they teach but different instructors fail very different fractions of students.  We can separate these effects in a set of classes by examining class-to-class variability of instructors who have taught multiple courses under CLASP10 grading.  Table \ref{tab4} shows each of these instructor's average and standard deviation, over classes, of the fraction of students with grades less than C$-$. Two new things show up in these data. First, the class-to-class standard deviation under CLASP10 varies by instructor but is relatively large for each of these instructors.  Instructors 5, 11, and 12 had the smallest variation with fail fractions varying by $\pm 2\%$ of their students and Instructor 10 had the largest variation over their classes where the fraction failing varies by $\pm 12\%$ of their students.  Note that the variation under CLASP4 never rises to 2\% for any instructor.  This shows that the class-to-class variation under CLASP10 is partly an \textbf{intra}-instructor effect.  Second, the fail rate averages also demonstrate a clear \textbf{inter}-instructor dependence of the class-to-class variation.  This instructor-dependent average varies from a minimum of 4\% of the students failing under Instructors 1 and 12 and a maximum of 25\% of the students failing under Instructor 10.  

\begin{table*}[htbp]
\caption{Eight instructors used the CLASP10 grade scale in at least 3 courses.  Instructors 1, 2, and 5 are also represented in Table \ref{tab3}.  N is the number of classes that the instructor taught under the specific grade scale over the time period of the study.  AvgFailRate is the average, over classes, of the fraction of students in a class who had grades less than C$-$ and StDevFailRate is the standard deviation, over classes, of the fraction of students in a class who had grades less than C$-$.  The variable $b_{LowGPA}$ is the linear coefficient from Equation \ref{eqn:FailvsGPAFrac}, with its standard error and P-value, for a regression limited to that instructor's classes. The relevant numbers from CLASP4 courses for the three instructors from Table \ref{tab3} are also shown.}
\label{tab4}
\begin{ruledtabular}
\begin{tabular}{c c c c c c c c}
\textbf{Instr} & \textbf{N} &\textbf{AvgFailRate} & \textbf{StDevFailRate} & $\mathbf{b_{LowGPA}}$ & \textbf{N} &  \textbf{AvgFailRate}
& \textbf{StDevFailRate}\\ 
 & \textbf{CLASP10} & \textbf{CLASP10} & \textbf{CLASP10} & \textbf{CLASP10} & \textbf{CLASP4} & \textbf{CLASP4}
& \textbf{CLASP4}\\
\hline
1 & 6 & 4 \% & 4 \% & 0.06 (0.30) P=0.86 & 14 & 0.6 \% & 0.7 \% \\
2 & 4 & 12 \% & 8 \% & 1.5 (0.4) P=0.07 & 6 & 1.6 \% & 1.1 \% \\
5 & 3 & 6 \% & 2 \% & -0.1 (0.9) P=0.92 & 14 & 1.1 \% & 0.8 \% \\
8 & 7 & 7 \% & 3 \% & 0.04 (0.22) P=0.87 \\
9 & 3 & 7 \% & 3 \% & 1.23 (0.11) P=0.06 \\
10 & 3 & 25 \% & 12 \% & -0.9 (0.8) P=0.45 \\
11 & 4 & 5 \% & 2 \% & 0.17 (0.18) P=0.44 \\
12 & 6 & 4 \% & 2 \% & -0.3 (0.2) P=0.37 \\
\end{tabular}
\end{ruledtabular}
\end{table*}

As a check to see if there is some regularity hidden in these numbers we have also used the model of Equation \ref{eqn:FailvsGPAFrac} for \textbf{each} instructor separately.  This models looks for a linear dependence of the class fail rate, $FracFail$, on that class' fraction of students with low incoming GPAs, $LowGPAFrac$, for each individual instructor. The linear coefficients, $b_{LowGPA}$, from the regression over each instructor's CLASP10 classes are shown in Table \ref{tab4} with the associated standard error and P-value.  Given the relatively large P-values, we find that \textbf{an instructor's fraction of students receiving grades $<$C$-$ has no clear dependence on the fraction of low GPA students in their class}.  Again the conclusions are robust under changing GPA cutoff.  These results suggest that variations in the students do not provide an easy explanation for the variations in the fail rates.  In summary, the CLASP10 grade scale results show both an intra-instructor increased variation and an inter-instructor increased variation in this low-grade part of the grade distribution when compared to the CLASP4 grade scale results.

\section{\label{sec:AppendixB}AGGREGATING GRADES - MEAN VS MEDIAN}

An alternative to using the mean to aggregate student grades is to use the median \cite{Brookhart2004}.  This is perhaps most straightforward with letter grades that are not attached to a numerical scale because they are explicitly ordinal but could also be argued to be appropriate for numerical grades that are referenced to letter grades when they are given by an instructor whose grading philosophy is that the grades they give are ordinal in nature.  For the CLASP courses the numerical grades are chosen from a continuous number line and were always averaged but these individual grades are ultimately connected to letter grades so one could argue that taking a median might be a reasonable choice here in constructing course grades.  In fact, the argument that it is best to consider even numerical grades to be ordinal and so use the median to aggregate grades has already been made by others \cite{Wright1994}.  For our grade data an interesting similarity between the two grade scales is that the overall median of all of the grades given to individual student answers under 4-point grading was 3.0 (exactly middle B) and the overall median under CLASP10 was 8.5 (also exactly a middle B).  It may help better understand the differences between the two grade scales to use the grade data we have discussed in this paper to compare, at the student level,  the two methods, median and average, of determining course grades.

The actual course grades given during the 10 years of these data are always made up of complicated weightings of the individual grading judgments.  This complicated weighting can even be student dependent because instructors would almost always drop each student's lowest quiz score, give the final exam more weight if the student performed better on that exam than on exams during the term, and adjust some student's grades slightly depending on their performance in their discussion/lab section.  This makes it impossible to devise a unique process of using the median function to determine a final grade that can be directly compared to the actual final grades.  Nevertheless, it may be useful in understanding the similarities and differences between the two grade scales to just construct a straight unweighted average of each student's individual grades and then compare those averages with a similarly unweighted median of each student's grades.  We find that 49 of the courses graded under 4-scale grading (11,708 students) and 24 courses under 10-scale grading (5,862 students) have a complete set of grades and so can be used in this way.

\begin{figure*}
\includegraphics[trim=1.5cm 5.2cm 1.5cm 4.2cm, clip=true,width=\linewidth]{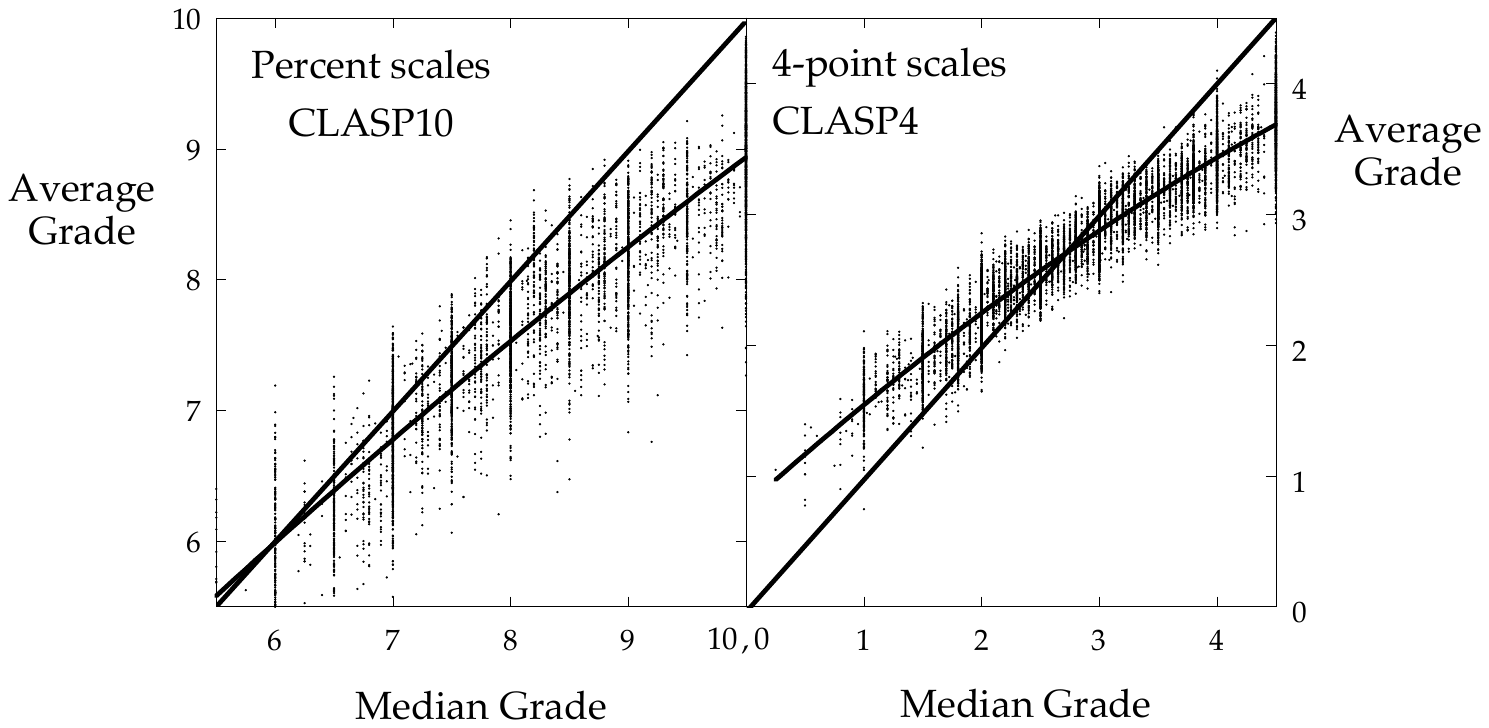}
\caption{The unweighted average of each student's set of grades is plotted against the unweighted median of these grades for each student from the classes for which all grade data were present.  Both grade scales are shown.  In each graph the straight line represents average = median and the curved line is a fit of all the grades to a quadratic polynomial done just to show the data trend.  The low grade region of the CLASP10 grade scale (the percent scale) is cutoff so that the A, B, C, and D ranges of the two scales occupy the same distance along each axis.}
\label{Fig8}
\end{figure*}

Figure \ref{Fig8} shows each student's unweighted average grade as a function of their unweighted median grade for both grade scales. The relationship between an average grade and a median grade, for a set of grades that have both an upper bound and a lower bound, is evident in the figure.  Students with many grades at the upper bound can have a median grade equal to the upper bound but any other grades they have will lower their average grade and exactly what other grades they have will determine how much lower their average grade is, so average grades will tend to be lower than median near the upper bound.  Conversely, average grades will tend to be higher than median when the median falls near the lower grade boundary. The result of those two general features is that average grades and median grades must be similar for some region in the middle of the grade space. For the grades given to the students in these courses, the average approximates the median in the B- to B region of grades under 4-point grading and at the F to D crossover region of grades under percent grading.  This means that under percent grading the region of grades that includes 92\% of student grades (median grade $\geq$ D) will tend to give lower grades when the average is used than when the median is used.

Another effect that is obvious from Figure \ref{Fig8} is that, for any specific value of the median grade, percent grading gives a much larger spread in average grades.  The standard deviation in the distribution of average grades is roughly twice as large for percent as for 4-point grading for any particular letter grade, determined by the median, from D$-$ to A$+$.  So, in the courses we are considering here, percent scale grading had both a larger systematic shift, of the average from the median, than 4-point grading and a larger random spread around that systematic shift.

\bibliography{mybibUSETHISONE }

\end{document}